\documentclass[pra, twocolumn, nofootinbib, amsmath, amssymb, superscriptaddress]{revtex4}
\usepackage[dvips]{graphics}
\usepackage{epsfig}

\begin{document}
\date{\today}
\title{Dynamics of Two Qubits: Decoherence and an Entanglement Optimization Protocol}
\author{C\'{e}sar A. Rodr\'{i}guez} 
\email[email: ]{carod@physics.utexas.edu}
\affiliation{  The University of Texas at Austin, Center for Complex Quantum Systems, 1 University Station C1602, Austin TX 78712}
\author{Anil Shaji}
\affiliation{ Department of Physics and Astronomy, University of New Mexico, Alburquerque NM 87131}
\author{E.~C.~G. Sudarshan}
\affiliation{  The University of Texas at Austin, Center for Complex Quantum Systems, 1 University Station C1602, Austin TX 78712}

\begin{abstract}
The evolution of two qubits coupled by a general nonlocal interaction is studied in two distinct regimes. In the first regime the purity of the individual qubits is interchanged through the entanglement shared by the two. We illustrate how this can be a mechanism for decoherence. In the second regime, the interaction entangles two initially pure qubits. The dynamical maps for the reduced unitary evolution of both initially simply separable and not-simply-separable states are found. We outline a protocol for optimizing the entanglement generation subject to constraints.
\end{abstract}


\maketitle 

\section{Introduction}

Universal, two qubit quantum gates are the basic units from which quantum information processing devices may be constructed \cite{{Deutsch95a},{Lloyd95a}}. Physical implementations of two qubit gates depend on understanding and controlling the possible interactions between the individual qubits. Zhang \emph{et. al.} \cite{Zhang03a} have obtained the most general form for non local two-qubit interactions. A complete picture of the dynamical possibilities allowed by this interaction is yet to be pieced together, although some numerical efforts have started to scratch the surface \cite{Munro01b}. Entanglement is a resource for a quantum information processor \cite{Nielsen00a,Bennet93a}, while entanglement of elements of the information processor with its environment leads to undesirable loss of purity and decoherence. Naturally, there has been a lot of interest\footnote{``As the strong man exults in his physical ability, delighting in such exercises as call his muscles into action, so glories the analyst in that moral activity which \emph{disentangles}." [sic] Poe, E.A. ``The Murders in the Rue Morgue" {\em Complete Stories and Poems of Edgar Allan Poe}. Ed. Doubleday \& Company, Inc. Garden City, NY (1966)} in understanding \cite{Vidal03a}, quantifying \cite{{Bennett96a},{Bennett96b},{Vedral97a},{Wootters98a},{Munro03a}}, and controlling \cite{{Haroche97a},{Haroche01a},{Monroe04a}} entanglement and also preventing decoherence. We show that both the loss of purity of qubits and the generation of entanglement between two qubits in an optimal way can be understood in terms of the dynamics generated by the same general two qubit interaction
under different conditions.

Our main result consists of an explicit expression for the dynamics of two qubits coupled by the general Hamiltonian. We calculate the dynamical map for both the initially simply separable and not simply separable two-qubit state. We focus on how the dynamics affect the purity of the individual qubits. We describe {\em purity swapping}, where one qubit is dynamically purified at the expense of another. This may be contrasted with entanglement swapping \cite{Vedral98a}. We also study the generation of entanglement by the dynamics. In \cite{Vidal01a} a method for maximizing the rate of entanglement generation has been proposed given some entangling Hamiltonian, and the capability to do instantaneous local operations on each qubit. We propose a protocol that optimizes any given entangling process subject to reasonable constraints that may be found in experiments.

The most general dynamics of the quantum state represented by a density matrix $\rho$ can be described in terms of a dynamical map  \cite{{Sudarshan61a},{Kraus71a}}:
\begin{equation}
\label{eq:DynMap} \rho\rightarrow\mathfrak{B}\left(\rho\right).
\end{equation}
The dynamical map represents the effect of the coupled unitary evolution of the system and its environment \cite{Sudarshan03a}. In other words:
\begin{equation}
\label{eq:TotalTrace} \rho^{A}\rightarrow \rho^{A}\left(t\right) = \mbox{Tr}_{B}\left[U \rho^{A}\otimes\rho^{B} U^\dag\right]=\mathfrak{B}\left(\rho^{A}\right),
\end{equation}
where $\rho^{A}$ is the state of the system and $\rho^{B}$ is that of the environment. In the case where the system and the environment are finite dimensional and start in a separable state, $\rho^{A}\otimes\rho^{B}$, $\mathfrak{B}$ is a matrix with positive eigenvalues and it is said to be completely positive.

\section{Evolution for two initially simply-separable qubits}

We concentrate on the simplest example, where both $\rho^A$ and $\rho^B$ are single qubits:
\begin{equation}
\label{eq:InitRho}
 \rho^A=\frac{\openone^A+\sum_i a_i\sigma^A_i}{2} , \ \rho^B=\frac{\openone^B+\sum_i b_i\sigma^B_i}{2} ,
 \end{equation}
 with $\sigma^A_i$ and $\sigma^B_i$ being the Pauli spin matrices for each of them. The Bloch vectors $\textbf{a}=(a_1,a_2,a_3)$ and $\textbf{b}=(b_1,b_2,b_3)$ provide a convenient way of parameterizing single qubit states that we will use in this paper. Together, $\rho^A$ and $\rho^B$ form the initially separable $4\times 4$ state, 
\begin{eqnarray}
\nonumber\label{eq:Ezero} E^{AB}(0) = \rho^{A}\otimes\rho^{B} = \frac{1}{4}\left[ \openone^A \otimes\openone^B + \right. \\   \sum_{i}\left(a_i\sigma^A_i\otimes\openone^B + b_i\openone^A\otimes\sigma^B_i\right)  + \sum_{i,j} \left. a_i  b_j\sigma^A_i\otimes\sigma^B_j\right],
\end{eqnarray}
where subscripts take values from $\{1,2,3\}$. The most general Hamiltonian for two qubits is:
\begin{equation}
\label{eq:GenHamil}
H=\sum_{i}\alpha_i\sigma_i^A\otimes\openone^B+\sum_{i}\beta_i\openone^A\otimes\sigma_i^B+\sum_{i,j}\Gamma_{ij}\sigma^A_i\otimes\sigma^B_j.
\end{equation}
In the interaction picture, it becomes, 
\begin{equation}
H\rightarrow \widetilde{H}(t)=\sum_{i,j}\widetilde{\Gamma}_{ij}(t)\sigma^A_i\otimes \sigma^B_j,
\end{equation}
which has nine parameters. Using local unitary transformations with three parameters each, the number of parameters can be brought down to three \cite{Glaser00a,Zhang03a,Cirac01a,Sudarshan03a}:
\begin{equation}
\label{eq:Hamil} \widetilde{H}(t)\rightarrow H(t)=\sum_i\gamma_i(t)\sigma^A_i\otimes\sigma^B_i.
\end{equation}

The time evolution of the overall state $E^{AB}$ is given by, $E^{AB}(t)=UE^{AB}(0)U^\dag$, where $U=\mathcal{T}e^{-i\int H(t)dt}$. For simplicity, we will assume that there is no free evolution for individual qubits, making $\gamma_i(t)\rightarrow\gamma_i$, and,
\begin{equation}
\label{eq:Unit} U = \prod_{j=1}^3\left[\cos(\gamma_j t)\openone^A\otimes\openone^B-i\sin(\gamma_j t )\sigma^A_j\otimes \sigma^B_j\right].
\end{equation}

To calculate $E^{AB}(t)$, we use the property that $\sigma^A_1\otimes\sigma^A_1, \sigma^B_2\otimes\sigma^B_2, \sigma^A_3\otimes\sigma^B_3$ all commute with each other. For each of the terms of $E^{AB}(0)$ we obtain:
\begin{eqnarray}\nonumber \label{eq:TermByTerm}
U\openone\otimes\openone U^\dag  & = & \openone \otimes\openone,
\\\nonumber U\sigma_i\otimes\sigma_i U^\dag  & = & \sigma_i  \otimes\sigma_i, 
\\\nonumber U\openone\otimes\sigma_iU^\dag &  = & \openone  \otimes\sigma_i e^{2it\left(\gamma_j\sigma_j\otimes\sigma_j+\gamma_k\sigma_k\otimes\sigma_k\right)},
\\ U\sigma_i\otimes\sigma_jU^\dag  & = & \sigma_i  \otimes\sigma_j e^{2it\left(\gamma_i\sigma_i\otimes\sigma_i+\gamma_j\sigma_j\otimes\sigma_j\right)},
\end{eqnarray} 
Using Eq. (\ref{eq:TermByTerm}), the evolution generated by Eq. (\ref{eq:Unit}) of the two qubit density matrix is: 
\begin{widetext} \begin{eqnarray}\nonumber \label{eq:Et} E^{AB}(t)& = &\frac{1}{4}\sum_{i=1}^3\left[\openone^A\otimes\openone^B +
a_i\left(C_j C_k\sigma^A_i\otimes\openone^B+S_j S_k\openone^A\otimes\sigma^B_i +C_k S_j\sigma^A_k\otimes\sigma^B_j-C_j S_k\sigma^A_j\otimes\sigma^B_k\right) \right.
\\\nonumber& & + b_i\left( C_j C_k\openone^A\otimes\sigma^B_i+S_j
S_k\sigma^A_i\otimes\openone^B  +C_k S_j\sigma^A_j\otimes\sigma^B_k-C_j S_k\sigma^A_k\otimes\sigma^B_j\right)+ a_i b_i \sigma^A_i\otimes\sigma^B_i
\\ \nonumber& & \left.  + a_i b_j\left(C_i C_j \sigma^A_i\otimes\sigma^B_j+S_i S_j\sigma^A_j\otimes\sigma^B_i  +C_i S_j\sigma^A_k\otimes\openone^B-C_j S_i \openone^A\otimes\sigma^B_k\right)\right.
\\ & & \left. + a_j b_i(C_i C_j \sigma^A_j\otimes\sigma^B_i+S_i S_j\sigma^A_i\otimes\sigma^B_j  +C_i S_j\openone^A\otimes\sigma^B_k-C_j S_i \sigma^A_k\otimes\openone^B)\right],
\end{eqnarray}\end{widetext}
where $C_i\equiv\cos\left(2t\gamma_i\right), S_i\equiv\sin\left(2t\gamma_i\right)$ and $i,j,k$ are cyclic. In other words, the coefficients of the Pauli matrices in Eq. (\ref{eq:Ezero}) transform as follows:

\begin{eqnarray} \nonumber\label{eq:Coeff} \left\{\begin{array}{c}   a_k \\   b_k \end{array}\right\} \rightarrow \left\{\begin{array}{c}   a_k \\   b_k \end{array}\right\}C_iC_j+\left\{\begin{array}{c}   b_k \\   a_k \end{array}\right\}S_iS_j \\+ \left\{\begin{array}{c}  a_i b_j\\   a_j b_i \end{array}\right\}C_iS_j -\left\{\begin{array}{c}  a_j  b_i\\   a_ib_j \end{array}\right\}C_jS_i, \end{eqnarray} where $\{i,j,k\}$ are cyclic, and \begin{eqnarray} \label{eq:Coeff2} a_i b_j & \rightarrow &   a_i b_j C_iC_j+ a_j b_i S_iS_j  \nonumber \\ && +  \epsilon_{ijk}(b_k C_jS_i -a_k C_iS_j), 
\end{eqnarray}
where it is not required for $\{i,j,k\}$ to be cyclic or distinct. 

\section{Reduced dynamics of two initially simply separable qubits}
To find the reduced dynamics of the system, $\rho^A$, we just need to carry out the partial trace from Eq. (\ref{eq:TotalTrace}). Using Eqs.~(\ref{eq:Et}), (\ref{eq:Coeff}), (\ref{eq:Coeff2}) and the fact that $\sigma^B_i$ are traceless, this is quite straightforward:

\begin{eqnarray} 
\label{eq:RhoT} 
\rho^{A}\left(t\right)  =  \frac{1}{2}\left(%
\begin{array}{cc}
  1+a_3(t) & a_1(t)-ia_2(t) \\
  a_1(t)+ia_2(t) & 1-a_3(t) \\
\end{array}%
\right),
\end{eqnarray}
where

\begin{eqnarray} \label{eq:RhoParams} 
\nonumber a_i\left(t\right)& =& a_i C_j
C_k+b_i S_j S_k\\&+&a_j b_k C_j S_k-a_k b_j C_k S_j.
\end{eqnarray}

The purity of the reduced density matrix changes with time:

\begin{eqnarray} 
\label{eq:Pt} 
\nonumber 
P^A(t)&=&\mbox{Tr}\left[\rho^{A}\left(t\right)^2\right]\\&=&\frac{1}{2}\left[1+a_1\left(t\right)^2+a_2\left(t\right)^2+a_3\left(t\right)^2\right].
\end{eqnarray}

Following a similar procedure we can find the evolution of $\rho^B(t)$ and its purity $P^B(t)$.
The time evolution of $\rho^A$ can be described using the dynamical map from Eq. (\ref{eq:DynMap}). The mapping matrix can be explicitly calculated to be \cite{Sudarshan61a}:\begin{widetext}
\begin{eqnarray}
\label{eq:DynMapB} \mathfrak{B}^{\left(0,t\right)} =\frac{1}{2}\left(%
\begin{array}{cccc}
  1+b_3S_1S_2+C_1 C_2  & b_2C_1S_2 -ib_1C_2S_1& \left(b_1S_3-b_2C_3\right)S_2 & \left(C_1+C_2\right) \\
  & & -i\left(b_2S_3+b_1C_3\right)S_1 & \times\left(C_3+ib_3S_3\right)\\
  b_2C_1S_3 +ib_1C_2S_1& 1+b_3S_1S_2  -C_1C_2& \left(C_2-C_1\right) & \left(b_1S_3+b_2C_3\right)S_2\\
  & & \times\left(C_3-ib_3S_3\right)& -i\left(b_2S_3-b_1C_3\right)S_1 \\
  \left(b_1S_3-b_2C_3\right)S_2 &  \left(C_2-C_1\right)&  &  \\
  +i\left(b_2S_3+b_1C_3\right)S_1 & \times\left(C_3+ib_3S_3\right) & 1-b_3S_1S_2-C_1C_2 & -b_2C_1S_2+ib_1C_2S_1\\
   \left(C_1+C_2\right)& \left(b_1S_3+b_2C_3\right)S_2 &  & \\
   \times\left(C_3-ib_3S_3\right) & +i\left(b_2S_3-b_1C_3\right)S_1 & -b_2C_1S_2-ib_1C_2S_1 & 1-b_3S_1S_2+C_1C_2 \\
\end{array}%
\right).
\end{eqnarray}
\end{widetext}

Note how the dynamical map carries the influence of $\rho^B$ through the parameters $b_1,\; b_2, \; b_3$. Thus, given any known initial state $\rho^A$ and sufficiently detailed evolution $\rho^A(t)$, its interaction with another unknown $\rho^B$ can be reconstructed, and even used to determine the parameters for the unknown state.

The most general map on a qubit can be implemented as the contraction of the unitary evolution of the given qubit coupled to at most two other qubits. We have restricted to the case where there is only one other qubit, thereby excluding certain maps. If $\rho^B$ is allowed to be a mixed state, the family of dynamical maps we are excluding by choosing a one qubit environment is very small \cite{Zalka02a}.

\section{Dynamical map of the general two qubit-state}
Eqs.~(\ref{eq:Coeff},\ref{eq:Coeff2}) may be extended in a straightforward fashion to the case where the initial state, $\tilde{E}$, is not simply separable, i.e.: \begin{eqnarray} 
\label{eq:EntZero}
\tilde{E}= \frac{1}{4}\sum_{i,j}\left[ \openone \otimes\openone +  e_{i0}\sigma_i\otimes\openone +  \right. \nonumber \\ \left. e_{0i}\openone\otimes\sigma_i  + e_{ij}\sigma_i\otimes\sigma_j\right],
\end{eqnarray}

where it is\emph{ not }necessary that $e_{ij} = e_{i0}\times e_{0j}$. This would include initially entangled states. The evolution for the state $\tilde{E}$ under Eq. (\ref{eq:Unit}) can be computed using the result from Eq. (\ref{eq:TermByTerm}): 
\begin{widetext} \begin{eqnarray}\nonumber \label{eq:Et} 
\tilde{E}(t)& = &\frac{1}{4}\sum_{i=1}^3\left[\openone^A\otimes\openone^B + e_{i0}\left(C_j C_k\sigma^A_i\otimes\openone^B+S_j S_k\openone^A\otimes\sigma^B_i +C_k S_j\sigma^A_k\otimes\sigma^B_j-C_j S_k\sigma^A_j\otimes\sigma^B_k\right) \right.
\\\nonumber& & + e_{0i}\left( C_j C_k\openone^A\otimes\sigma^B_i+S_j
S_k\sigma^A_i\otimes\openone^B  +C_k S_j\sigma^A_j\otimes\sigma^B_k-C_j S_k\sigma^A_k\otimes\sigma^B_j\right)+ e_{ii} \sigma^A_i\otimes\sigma^B_i
\\ \nonumber& & \left.  + e_{ij}\left(C_i C_j \sigma^A_i\otimes\sigma^B_j+S_i S_j\sigma^A_j\otimes\sigma^B_i  +C_i S_j\sigma^A_k\otimes\openone^B-C_j S_i \openone^A\otimes\sigma^B_k\right)\right.
\\ & & \left. + e_{ji}(C_i C_j \sigma^A_j\otimes\sigma^B_i+S_i S_j\sigma^A_i\otimes\sigma^B_j  +C_i S_j\openone^A\otimes\sigma^B_k-C_j S_i \sigma^A_k\otimes\openone^B)\right].
\end{eqnarray}\end{widetext}

Again, to find the reduced dynamics of the system we just need to carry out the partial trace, and the evolution of each of the components become:

\begin{eqnarray} \label{eq:RhoParamsEnt} 
\nonumber e_{i0}\left(t\right)& =& e_{i0} C_j C_k+e_{0i} S_j S_k\\&+&e_{jk} C_j S_k-e_{kj} C_k S_j.
\end{eqnarray}

As before, we would like to construct the dynamical map for this evolution. This time, the map has to carry the parameters of the trace out qubit \emph{as well as the cross-terms}. Some of the terms that could be factored before into parameters in the reduced state and parameters in the map cannot be. This leads to the following map:
\begin{widetext}
\begin{eqnarray}
\label{eq:DynMapBEnt} \mathfrak{\widetilde{B}}^{\left(0,t\right)} =\frac{1}{2}\left(%
\begin{array}{cccc}
  1+C_1 C_2        &   & e_{01}S_2 S_3 -i e_{02}S_3 S_1 & \\
     +e_{03}S_1S_2 & 0& +e_{23}C_2 S_3 - e_{32} C_3 S_2 & \left(C_1+C_2\right)C_3 \\
  +e_{12}C_1 S_2-e_{21} C_2 S_1 &   & -i e_{31}C_3 S_1 +i e_{13}C_1 S_3      & \\
  
  & 1-C_1 C_2 &                                            & e_{01}S_2 S_3 -i e_{02}S_3 S_1\\
 0 & +e_{03}S_1S_2& \left(C_2-C_1\right)C_3 & +e_{23}C_2 S_3 - e_{32} C_3 S_2 \\
  & +e_{12}C_1 S_2-e_{21} C_2 S_1 & &  -i e_{31}C_3 S_1 +i e_{13}C_1 S_3  \\
  
  e_{01}S_2 S_3 +i e_{02}S_3 S_1 & & 1-C_1 C_2& \\
   +e_{23}C_2 S_3 - e_{32} C_3 S_2 &  \left(C_2-C_1\right)C_3 &    -e_{03}S_1S_2 &  0\\
   +i e_{31}C_3 S_1 -i e_{13}C_1 S_3   &   & -e_{12}C_1 S_2+e_{21} C_2 S_1  &  \\
  
  & e_{01}S_2 S_3 +i e_{02}S_3 S_1 & & 1+C_1 C_2  \\
  \left(C_1+C_2\right)C_3 &  +e_{23}C_2 S_3 - e_{32} C_3 S_2   & 0 &    -e_{03}S_1S_2 \\
                          &  +i e_{31}C_3 S_1 -i e_{13}C_1 S_3    &  &  -e_{12}C_1 S_2+e_{21} C_2 S_1 \\
\end{array}%
\right).
\end{eqnarray}
\end{widetext}

The first thing to note about this map is that there are some elements that are $0$. This is due to the fact that not all terms can be decomposed as a product and split between the state and the action of the map. In other words, the map $\widetilde{\mathfrak{B}}$ carries all the information in the cross terms of the initial two qubit state. To guarantee that this map correspond to something physical, we need to observe that due to the initial correlations of the bipartite state, only certain values are permitted for the reduced intimal state. The map has some information of its allowed domain, and as long as it acts on it the evolution can still be given physical interpretation \cite{Shaji05a} and its eigenvalues in general can be negative. There are some experimental examples of these non-completely positive maps \cite{Howard05a} and also their role in decoherence control techniques \cite{Terno05a}.

\section{Decoherence modelling}

Now, we go back to the case of initially simply-separable states. We want to study certain kinds of evolution to show the use of these maps in some decoherence modelling schemes.

\begin{figure}[htb]  \begin{center}      \includegraphics[width=.9\columnwidth]{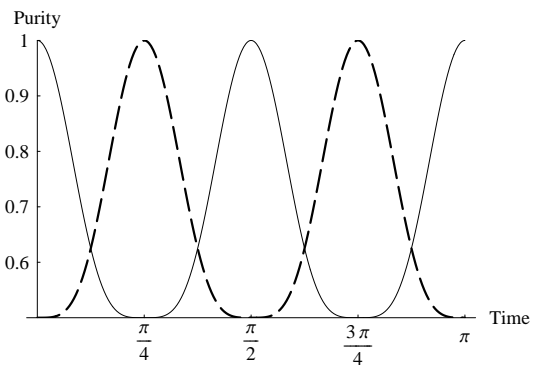}    \end{center}    \caption{Purity as a function of time for $a_1= \gamma_2=\gamma_3=1,a_2=a_3=b_1=b_2=b_3=\gamma_1=0$. The solid line represents $P^A(t)$ while the dashed line represents $P^B(t)$. At $t=\pi/4$ purity has been totally swapped.} \label{fig1} \end{figure}

 Consider the case where $\rho^A(0)$ is pure while $\rho^B(0)$ is fully mixed. Assume some interaction where the only non-zero parameters are $a_1= \gamma_1=\gamma_2=\gamma_3=1$. This interaction is chosen to be of the "swap-gate" form, but many interactions can have a similar effect. Figure \ref{fig1} shows how the purity of $\rho^A(t)$ and the purity of $\rho^B(t)$ change with time.  As time progresses, they start to get entangled and exchange purity through entanglement. $P^A(t)$ and $P^B(t)$ are equal at $t=\pi/8$, where some purity has been lost to entanglement. At $t=\pi/4$ they become separable again \cite{Peres96a} purifying $\rho^B$ at the expense of $\rho^A$, a dynamical process we call \textit{purity swapping}. If only a certain component of the qubit is measured and no correlations studied, this might look like Rabi oscillations, although the Bloch vector is \emph{not} rotating, but oscillating in length.

We remark that for a weak coupling of this type, at very short times, $P^A(\delta t)$ can only decrease (except in trivial cases). This can be considered to be a mechanism of decoherence. We can model a reservoir as a stream of $\left\{\rho^B_i\right\}$, where each of them interact independently for a short average time \cite{Rau63a}, swapping some purity from $\rho^A$ to each $\rho^B_i$, but stopping the coupling before there is enough time to return the stolen purity. We can think of this as a quantum version of the Boltzmann gas \cite{Merkli05a}. This corresponds to acting with the dynamical map from Eq. (\ref{eq:DynMapB}) in sequence: \begin{equation} \label{eq:Decoh} \rho^A\rightarrow \mathfrak{B}^{\left(t_{n-1},t_n\right)}\circ\mathfrak{B}^{\left(t_{n-2},t_{n-1}\right)}\circ\ldots\circ\mathfrak{B}^{\left(t_0,t_1\right)}\left(\rho^A\right). \end{equation} By controlling the strength, duration and number of these reservoir interactions it is possible to model decoherence processes, as well as desired, using only a finite number of degrees of freedom for the reservoir. 

There are many cases where the periodic couplings are a good approximation and its properties exploited to prevent decoherence. Bang-bang control, for example, is a technique where the periodicity (on average) of a decoherence coupling can be synchronized with a local control pulse effectively creating a decoherence-free space for the state of interest  \cite{Viola99a,Uchiyama06a}.

\section{Optimal entanglement generation}

Another interesting regime to study is related to the creation of maximally entangled Bell states. Assume that we have initially two pure states, $\{a_1=1,\;a_2=a_3=0\}, \{b_2=1,\;b_1=b_3=0\}$. At a specific time $t_{bell}$, each of the qubits' purity goes to a minimum. For $\gamma_3=1$, the minimum is at $t_{bell}=\pi/4$, and we get that $E^{AB}(t_{bell})$ is: \begin{eqnarray}\label{eq:FullEnta}\ \frac{1}{4}\left[\openone^A\otimes\openone^B+\sigma_1^A\otimes\sigma_2^B  -\sigma^A_2\otimes\sigma^B_3+\sigma^A_3\otimes\sigma^B_1\right],
\end{eqnarray}
that, given some freedom to choose the basis for $\rho^B$, would be equivalent to the Bell state $\frac{1}{\sqrt{2}}\left(|00\rangle+|11\rangle\right)$. A similar, experimentally feasible time-reversed procedure would be responsible for extracting purity out of entanglement \cite{Schrodinger35a}.

How much can an entanglement creation experiment be improved? From Eq. (\ref{eq:Et}) an entanglement optimization protocol can be determined, given certain reasonable constraints. In particular, we would like to show how given a certain two-qubit state and certain experimental limitations in the Hamiltonian that can be applied, the evolution can be optimized to maximize the entanglement between them. 

Lets illustrate this with a simple example for the case of the creation of entanglement from initially pure qubits. Limitations on the allowed interaction and their duration would come from particular experimental requirements. This example, although very simple, contains all the elements to illustrate how a procedure like this can be implemented. The more general protocol is discussed afterwards.

Imagine some experimental setup that prepares two qubits with the Bloch vectors $\textbf{a}$ and $\textbf{b}$ representing their states oriented at some angle with respect to each other. The aim is to entangle these two. Assume that only two kinds of couplings between them are allowed and we can only act first with one kind followed by the other, due to some experimental constraints. The only thing that can be controlled is the duration for which each coupling is used. The total time available is restricted by decoherence. This type of interaction can be similar to the "strongly modulating" pulses in nuclear magnetic resonance implementation of quantum algorithms, but is not limited to it. How long should we act with each of them to maximize the entanglement rate? 

Let the two qubits $\rho^A$ and $\rho^B$, initially be in the states given by $a_1=1$, $a_2=a_3=0$ and $b_1=b_2=1/\sqrt{2}$, $b_3=0$. The allowed interactions are $\gamma_2=1,\;\gamma_1=\gamma_3=0$ for some time $[0,t^\prime]$ and followed by $\gamma_3=1,\;\gamma_1=\gamma_2=0$ at $t^\prime$ for an interval $(t^\prime,\pi]$. Using Eq. (\ref{eq:Et}) we can calculate the state $E^{AB}(t)$ of the system at $t^\prime$. Using $E^{AB}(t^\prime)$ as the new initial condition in Eq. (\ref{eq:EntZero}) and the new coupling, $\gamma_3=1$ we can compute the state of the two qubits during $(t^\prime,\pi]$. Eq. (\ref{eq:RhoT}) gives us the reduced density matrix of one of the qubits as a function of time from which we can compute its purity at every time.

What is the time $t^\prime$ that gives us the maximum entanglement? This protocol is not dependent on a particular measure of entanglement, but we will choose for simplicity the entropy of entanglement \cite{Bennett96a}, which is a good measure as long as $E^{AB}(t)$ remains pure. Since it is monotonically related to the linear entropy \cite{{Vedral00a},{Cirac01a}}, the entanglement measure $\mathcal{E}$ can be also chosen:
\begin{equation}
\label{eq:EMeas}
\mathcal{E}\sim1-\mbox{Tr}\left[(\rho^A)^2\right]=1-\mbox{Tr}\left[(\rho^B)^2\right]=1-P.
\end{equation}
Using Eqs.~(\ref{eq:RhoParams}) and (\ref{eq:Pt}), we find that for our choices of the parameters:
\begin{equation} \label{eq:Pexamp} \mathcal{E}\sim1-\frac{3\cos\left(8t^\prime\right)+29}{32}.
\end{equation}
which is maximized at $t^\prime=\pi/8,\;3\pi/8,\;5\pi/8$ or $7\pi/8$. We admit that this example is rather simple, and was chosen because it can be easily solved algebraically. More complicated and realistic couplings and limitations can be solved numerically with similar results.

This method can be applied to any interaction of the form Eq.~(\ref{eq:Hamil}), allowing us to optimize a very general class of entanglement creation procedures. It is significantly different from the procedure proposed in \cite{Vidal01a}, in that we do not need to assume full control over the local transformations on each qubit at all times.

The general scheme of this entanglement optimization protocol can be summarized as follows:
\begin{enumerate}
    \item Choose the initial conditions for the system, and express them in the form of either Eq.~(\ref{eq:Ezero}) or Eq.~ (\ref{eq:EntZero}).
    \item Identify the coupling parameters and constraints for Eq.~(\ref{eq:Hamil}). Compute the evolution due to the unitary operator from Eq.~(\ref{eq:Unit}) by using the transformations in Eqs.~(\ref{eq:Coeff},\ref{eq:Coeff2},\ref{eq:Et}).
    \item Choose the measure of entanglement of your preference \cite{Bennett96a,Bennett96b,Vedral97a,Wootters98a,Munro03a}.
    \item Considering the experimental constraints, optimize with respect to the desired parameters.
    \end{enumerate}

\section{Conclusions}

In summary, we have explicitly calculated the evolution generated  by a general interaction between two qubits, opening the doors for studying all sorts of entangling interactions and experimentally realizable universal quantum gates. We compute  the Dynamical Map of the most general evolution for two qubits, and showed the different nature that the Map can have when the state is initially simply-separable of not. Through examples, we illustrated how purity and entanglement are interchangeable quantities. We studied dynamical purity swapping, and its connection to decoherence phenomena. Our simple two qubit interaction is flexible enough to study the fundamentals of decoherence that are usually studied as an interaction with an infinite degrees of freedom reservoir. We described a procedure to generate Bell States from pure states, and its converse, purification. Finally, we proposed a practical protocol to optimize a given entangling procedure under realistic constrains.

The authors would like to thank Thomas F. Jordan for several insightful discussions. One of the authors (C.A.R.) would like to thank Kavan Modi and Mark Selover for proofreading the manuscript. A.~S.~Acknowledges the support of US Office of Naval Research Contract No. N00014-03-1-0426.

\bibliography{p2005}

\end{document}